\documentstyle[preprint,aps]{revtex}
\begin{document}
\newcommand{\be}{\begin{equation}}
\newcommand{\ee}{\end{equation}} 
\newcommand{\lb}{\label}
\newcommand{\en}{\epsilon}
\newcommand{\ven}{\varepsilon}
\newcommand{\bu}{{\bf u}}
\newcommand{\bx}{{\bf x}}
\newcommand{\bk}{{\bf k}}
\newcommand{\bh}{{\bf h}}
\newcommand{\bF}{{\bf f}}
\newcommand{\bJ}{{\bf J}}
\newcommand{\fL}{\overline{{\bf f}}}
\newcommand{\uL}{\overline{{\bf u}}}
\newcommand{\us}{{\bf u}^{\prime}}
\newcommand{\oL}{\overline{{{\mbox{\boldmath $\omega$}}}}}
\newcommand{\os}{{{{\mbox{\boldmath $\omega$}}}}^{\prime}}
\newcommand{\pL}{\overline{p}}
\newcommand{\btau}{{\mbox{\boldmath $\tau$}}}
\newcommand{\bdot}{{\mbox{\boldmath $\cdot$}}}
\newcommand{\btimes}{{\mbox{\boldmath $\times$}}}
\newcommand{\grad}{{\mbox{\boldmath $\nabla$}}}
\newcommand{\bom}{{\mbox{\boldmath $\omega$}}}

\pagestyle{myheadings}
\draft
\preprint{Submitted to Phys. Rev. Lett.}

\title{Intermittency in the Joint Cascade of Energy and Helicity}
\author{Qiaoning Chen$^{1}$, Shiyi Chen$^{1,2,3}$, Gregory L. Eyink$^{4}$ 
and Darryl D. Holm$^{2}$}
\address{
${}^{1}$Department of Mechanical Engineering, The Johns Hopkins University,
 Baltimore, MD 21218\\
${}^{2}$Center for Nonlinear Studies and Theoretical Division, Los Alamos 
National Laboratory, Los Alamos, NM 87545\\
${}^{3}$National Key Laboratory for Turbulence Research, 
Peking University, China\\
${}^{4}$Department of Mathematics, University of Arizona, Tucson, AZ 85721\\
}
\maketitle

\begin{abstract}
The statistics of the energy and helicity fluxes in isotropic turbulence are 
studied using high resolution direct numerical simulation. The scaling exponents of the energy flux agree with those of the transverse velocity structure 
functions through refined similarity hypothesis, consistent with Kraichnan's
 prediction \cite{Kr74}. The helicity flux is even more intermittent than the
 energy flux and its scaling exponents are closer to those of the passive 
scalar. Using Waleffe's helical decomposition, we demonstrate that the 
existence of positive mean helicity flux inhibits the energy transfer in the 
negative helical modes, a non-passive effect.              

\end{abstract}


The classical theories of fully-developed turbulence \cite{K41} were dominated 
by the concept of the energy cascade to small scales. However, kinetic energy is not the only local conserved integral of the inviscid equations of motion, the three-dimensional (3D) incompressible Euler equations. Since the classical 
theories were developed, it was discovered \cite{Mor,Mof} that there is a 
second quadratic invariant, the {\it helicity}:
\be       H(t) = \int d\bx \,\,\bu(\bx,t)\bdot\bom(\bx,t).       \lb{1} \ee
Here $\bu$ is the velocity field and $\bom=\grad\btimes\bu$ the vorticity field. Nonzero mean values of the helicity are now known to occur naturally in a wide variety of geophysical flows, such as hurricanes and tornadoes \cite{LET}. It was proposed in \cite{BFLLM,Kr73} that, if the large-scales of the flow are helical (parity non-invariant), then there should be a {\it joint cascade} of both energy and helicity to small-scales. In that case, the helicity spectrum as well as the energy spectrum should satisfy a $-5/3$ law in the inertial range:
$H(k) \sim C_H (\delta/\epsilon^{1/3}) k^{-5/3}.$  Just as for a passive scalar, the spectrum of helicity was predicted to be linearly proportional to its mean flux $\delta$. This predicted joint cascade was confirmed recently in a numerical simulation of 3D Navier-Stokes \cite{BO97}, which seems to support the idea of a relatively minor role for helicity in hydrodynamic turbulence \cite{Kr73}.
 \cite{OG,VEI} give theoretical support to the joint cascade picture. However, 
helicity cannot act exactly as a passive scalar. It is well-known that presence of helicity suppresses the nonlinear transfer and acts to inhibit the energy 
cascade \cite{Kr73,PS,HK}. In 3D magnetohydrodynamic turbulence, the presence of mean magnetic helicity may affect the inertial-range scaling laws \cite{MB}.
 Perhaps the most intriguing evidence of a hidden role for helicity in 
hydrodynamic turbulence is from the so-called ``shell models''. In a family of
 such models, the GOY shell models, it has been found numerically in \cite{KLWB} that the scaling exponents of the energy flux are nearly identical to those for 3D Navier-Stokes precisely for the members of the family which have a 
helicity-like invariant. The statistics of the helicity flux itself have also
 been studied in the GOY models \cite{DG} and in a related class of 
``helical shell models'' \cite{BPT}. However, so far the statistics 
of the helicity flux have yet to be explored in 3D turbulence. It is the purpose of this Letter to study the statistics of energy and helicity fluxes in 
3D hydrodynamical turbulence by direct numerical simulations, both with and
 without a nonzero mean helicity.

We have simulated the Navier-Stokes (NS) equation in a $512^3$ domain. 
The kinetic energy is forced in the first two shells \cite{CS}. To add positive mean helicity into the flow, we rotate the real and imaginary parts of the 
velocity vector Fourier amplitude also in the first two shells to be always
 perpendicular to each other with the same handedness \cite{PS}. The NS 
equation was solved using a pseudo-spectral parallel code with full dealiasing. The time stepping was a second-order Adam-Bashforth method with a modified Euler method for the initial time step. A statistical stationary state was achieved
 after five large-eddy turnover times. In Fig.~1 we plot the energy and helicity spectra of this final steady-state, in the case with mean helicity input. Both spectra have about a decade and a half where a $-5/3$ power-law holds. In Fig.~2 we show for the same simulation the mean spectral fluxes of energy and helicity as a function of wavenumber, normalized by mean energy dissipation 
$\varepsilon=\nu\langle|\grad\bu|^2\rangle$ and mean helicity dissipation 
$\delta= 2\nu\langle\grad\bu:\grad\bom\rangle.$  There is about a decade of 
inertial range where these fluxes are constant.  

The importance of a {\it local energy flux} for studying intermittency in the 3D energy cascade was first emphasized by Kraichnan \cite{Kr74}, who used banded 
Fourier series to define such a quantity. The local flux $\Pi_{E,\Delta}(\bx)$ 
measures transfer of energy into small length scales $<\Delta$ at a fixed point $\bx$ in physical space. Kraichnan proposed a {\it refined similarity 
hypothesis} (RSH) which relates the scaling exponents $\zeta_p$ of the velocity structure functions, defined as $\langle|\Delta_\ell \bu|^p\rangle\sim\ell^{\zeta_p}$, to the scaling exponents $\tau_p$ of the energy flux, defined by
 $\langle|\Pi_{E,\Delta}|^p\rangle \sim\Delta^{\tau_p}$. 
Here $|\Delta_\ell{\bf u}|$ is the magnitude of the vector velocity increment.
Precisely, the RSH relation is $\zeta_p={{p}\over{3}}+\tau_{p/3}.$ Equivalently, this relation may be stated as the equality $\zeta_p=\zeta_p^E$, with the 
latter defined by 
\be \langle|\Delta\cdot\Pi_{E,\Delta}|^{p/3}\rangle \sim \Delta^{\zeta^E_p} \lb{2} \ee
for $L\gg\Delta\gg \eta$, where $L$ is the integral length and $\eta$ the dissipation length. To test this relation here we shall use instead a smooth filter to differentiate the large-scale and small-scale modes, as in our earlier 
work \cite{Ey95}. This is the same method used in the large-eddy simulation 
(LES) modeling scheme \cite{MK}. A low-pass filtered velocity 
$\uL=G_\Delta*\bu$ with scales $<\Delta$ removed obeys the equation
\be \partial_t\uL+ (\uL\bdot\grad)\uL = \fL-\grad\pL -\grad\bdot\btau \lb{3} \ee
in the limit of high Reynolds number, where viscous terms can be neglected. Here $\fL,\pL$ are the filtered forcing and pressure, respectively, and $\btau=\overline{\bu\otimes\bu}-\uL\otimes\uL$ is the {\it turbulent stress}, or spatial 
momentum transport induced by the eliminated small-scale turbulence. From this
 equation, a balance equation is easily derived for the local density 
$e_\Delta ={{1}\over{2}}|\uL|^2$ of the large-scale energy \cite{Ey95}:
\be \partial_t e_\Delta(\bx,t)+\grad\bdot\bJ_{E,\Delta}(\bx,t)= F_{E,\Delta}(\bx,t)
                                                         -\Pi_{E,\Delta}(\bx,t) \lb{4} \ee
in which the current $\bJ_{E,\Delta}$ represents space-transport of the large-scale energy, $F_{E,\Delta} = \fL\bdot\uL$ is the energy input from the force, and  
\be \Pi_{E,\Delta}(\bx,t) = -\grad\uL(\bx,t):\btau(\bx,t) \lb{5} \ee
is the {\it energy flux} out of the large-scales and into the small-scale modes. See also \cite{BO98,CM}. 

In the same way we can derive a balance equation for the density $h_\Delta=\uL\bdot\oL$ of the large-scale helicity:
\be \partial_t h_\Delta(\bx,t)+\grad\bdot\bJ_{H,\Delta}(\bx,t)=F_{H,\Delta}(\bx,t)-\Pi_{H,\Delta}(\bx,t) \lb{6} \ee
See \cite{CCED}. Here $\bJ_{H,\Delta}$ is a space transport of large-scale helicity and $F_{H,\Delta}$ is the input from the forcing, while 
\be \Pi_{H,\Delta}(\bx,t) = -2\grad\oL(\bx,t):\btau(\bx,t) \lb{7} \ee 
is the local {\it helicity flux} to small scales $<\Delta$. A set of scaling exponents $\zeta_p^H$ corresponding to the helicity cascade can be defined by 
\be \langle|\Delta^2\Pi_{H,\Delta}|^{p/3}\rangle \sim \Delta^{\zeta^H_p} \lb{8} \ee
for $L\gg\Delta\gg \eta$. 

In Fig.~3 we plot the structure functions of the energy and helicity fluxes that appear on the lefthand sides of Eqs.(\ref{2}),(\ref{8}) for integer values of $p$ from 1 to 9. 

We use the Extended Self-Similarity (ESS) procedure \cite{ESS} to extract the scaling exponents $\zeta^E_p$ and $\zeta^H_p$ for $30 \leq \Delta/\eta \leq 80$. 
The results are shown in Fig.~4. Together with these we plot the scaling 
exponents $\zeta_p^T$ of the transverse velocity structure functions. 
The transverse velocity differences are known to be more intermittent than 
the longitudinal ones \cite{CSNC,RA} and thus must dominate in the structure 
function for which Kraichnan's RSH was proposed, which includes {\it all} velocity components. As may be seen from Fig.~4, the scaling exponents $\zeta^T_p$ and $\zeta^E_p$ are quite close for each $p$ , in agreement with Kraichnan's RSH 
\cite{Kr74,Ey95}. However, the scaling exponents for helicity flux are smaller, $\zeta_p^H<\zeta_p^E,$ indicating that the helicity flux is intrinsically {\it more intermittent} than the energy flux. This is consistent with the picture of
 the helicity acting similarly as a passive scalar, since it is well-known that the scaling exponents of the scalar are smaller than those of the advecting 
velocity itself \cite{sreeni,Shiyi}. As shown in the inset of  Fig.~4, the 
scaling exponents of helicity flux are indeed rather close to those of passive
 scalars \cite{CAO}. It is worth emphasizing that the relation between the 
scaling exponents of energy flux and helicity flux found here is exactly the
 opposite of that observed in the shell models. In \cite{DG} it was shown
 numerically that the helicity flux in the GOY shell model is
 {\it less} intermittent than the energy flux there, while in \cite{BPT} it
 was shown for the ``helical GOY3 model'' that energy and helicity fluxes 
are {\it equally} intermittent. Thus, despite the fact that the energy flux 
statistics of the helicity-conserving shell models are very similar to those of 3D Navier-Stokes \cite{KLWB}, nevertheless the helicity-flux statistics of these shell models are qualitatively different from those of Navier-Stokes. This is
 discussed further in \cite{CCED}. 

Fluctuations of the fluxes $\Pi_{E,\Delta}(\bx,t)$ and $\Pi_{H,\Delta}(\bx,t)$ in the joint cascade can also be described by the single-point probability 
density functions (PDF's), which we have calculated at various values 
of $\Delta$ in the inertial range. In Fig.~5 are plotted these PDF's for
 $\Delta/\eta=64$. $\mu$ represents the mean of the flux and $\sigma$ stands for the standard deviation of the flux. The PDF of the energy flux agrees with the results reported earlier in \cite{CM}. There is an obvious skewness with a long tail to the right, indicating the forward cascade of energy preferentially to
 smaller scales. To our knowledge, the result on the helicity flux PDF
 in Fig.~5 is entirely new. In contrast to the PDF of the energy flux, it is 
nearly symmetric. This is because the helicity is indefinite in sign and can be both positive and negative, while kinetic energy is always positive. Therefore, the long tail of helicity flux to the left does not indicate backward transfer
 of positive helicity but instead forward transfer of negative helicity. This 
argument can be made precise by means of the {\it helical decomposition} of the velocity field \cite{Waleffe}. One sets $\bu=\bu^+ +\bu^-$ with
\be \bu^\pm(\bx,t) = \sum_\bk a_\pm(t) \bh_\pm(\bk)e^{i\bk\bdot\bx} \lb{9} \ee
where $\bh_s(\bk)$ for $s=\pm$ are eigenvectors of the curl: $i\bk\btimes\bh_s(\bk)=s|\bk|\bh_s(\bk).$ As shown in \cite{Waleffe}, the $+$ modes carry only positive helicity and the $-$ modes only negative helicity. If we likewise decompose vorticity as $\bom=\bom^+ + \bom^-$, we can define
\begin{eqnarray}
    \Pi_{E,\Delta}^\pm(\bx,t) = -\grad\uL^\pm(\bx,t):\btau(\bx,t), \cr 
    \Pi_{H,\Delta}^\pm(\bx,t) = -2\grad\oL^\pm(\bx,t):\btau(\bx,t) \lb{10} 
\end{eqnarray}
and then  $\Pi_{E,\Delta}^s$ represents flux of energy from the large-scale $s$-modes into the small scales and $\Pi_{H,\Delta}^s$ represents the corresponding quantity for helicity, with $s=\pm$. See \cite{CCED}. 

In Fig.~6(a) are plotted the 1-point PDF's of $\Pi_{H,\Delta}^\pm$ for nonzero mean helicity with the inset for zero mean helicity. It may be seen that the PDF of the $+$ mode is skewed to the right, while that of the $-$ mode is skewed to the left, indicating that both positive and negative helicity prefer to cascade forward to smaller scales. We can not see much difference between nonzero mean
 helicity and zero mean helicity. In Fig.~6(b) are plotted the 1-point PDF's of $\Pi_{E,\Delta}^\pm$ for nonzero mean helicity with the inset for zero mean
 helicity. Although the PDF of $\Pi_{E,\Delta}^+$ is similar to that for the
 total energy flux itself, and skewed to the right, the PDF 
of $\Pi_{E,\Delta}^-$ is much more nearly symmetrical for nonzero mean helicity case . Therefore, the positive mean flux of helicity seems to inhibit forward
 transfer of energy in the $-$ modes, exhibiting the ``non-passive'' character
 of helicity in fully-developed turbulence. This discovery is in agreement with the assertion of Kraichnan \cite{Kr73} that interactions will be strongest 
between the helical modes of the opposite signs. See also \cite{Waleffe}.  

To see further how the helicity flux affects the energy flux we have calculated 
the PDF of the energy flux conditioned on the value of the helicity flux,
in the case with $\delta\neq 0$. In Fig.~7 are shown two PDF's of energy 
flux, one conditioned on helicity flux in the range [0.0, $3\sigma^H$] and the 
other in the range [$3\sigma^H$, $6\sigma^H$]. We can see that energy flux is 
more intermittent when the helicity flux is small than when the helicity 
flux is large. Clearly, the tail of the PDF with the large helicity flux
is reduced compared to that with the small helicity flux, implying that
large helicity flux ``blocks'' large energy flux. However, the effect 
of the conditioning upon helicity flux is not entirely simple. While very 
large energy flux events are suppressed in the presence of large helicity
flux, the probability of larger fluctuations near the center of the distribution
is actually increased. It is an interesting question whether similar effects 
would be seen on energy flux when conditioning on values of a passive scalar
flux. The effects observed imply only statistical correlations between two
fluxes and no direct dynamical interaction. A detailed theory of the 
phenomena remains to be developed.

{\bf Acknowledgments.}

We thank Robert M. Kerr, Charles Meneveau, Mark Nelkin, Xin Wang and Victor 
Yakhot for helpful discussions. Simulations are performed at the Advanced Computing Laboratory at Los Alamos National Laboratory and the cluster computer 
supported by the NSF grant CTS-0079674 at the Johns Hopkins University.


\pagebreak
\vspace{0.2in}
\noindent {\bf Figure Captions}
\begin{description}
\item{
FIG.~1. Energy and helicity spectra.
}
 
\item{
FIG.~2. Normalized energy and helicity fluxes versus wavenumber.
}
 
\item{
FIG.~3. Structure functions (a) of energy flux, (b) of helicity flux. The scaling exponents are measured using the data between two dashed lines.}
 
\item{
FIG.~4. Scaling exponents of transverse velocity increments, energy flux, and helicity flux. In the inset we show the scaling exponents of helicity
flux($\bullet$) and passive scalar($\Diamond$)}
 
\item{
FIG.~5. 1-point PDF's of energy and helicity fluxes.
}
 
\item{
FIG.~6. 1-point PDF's of (a) $\pm$ helicity fluxes, (b) $\pm$ energy fluxes for
different mean helicity inputs $\delta$ when $\Delta/\eta=64$. The inset is for
$\delta=0.0$. }
 
\item{
FIG.~7. Normalized energy flux PDF's conditioned on helicity flux when
$\Delta/\eta=64$, $\delta\neq 0$.
}
 
\end{description}       
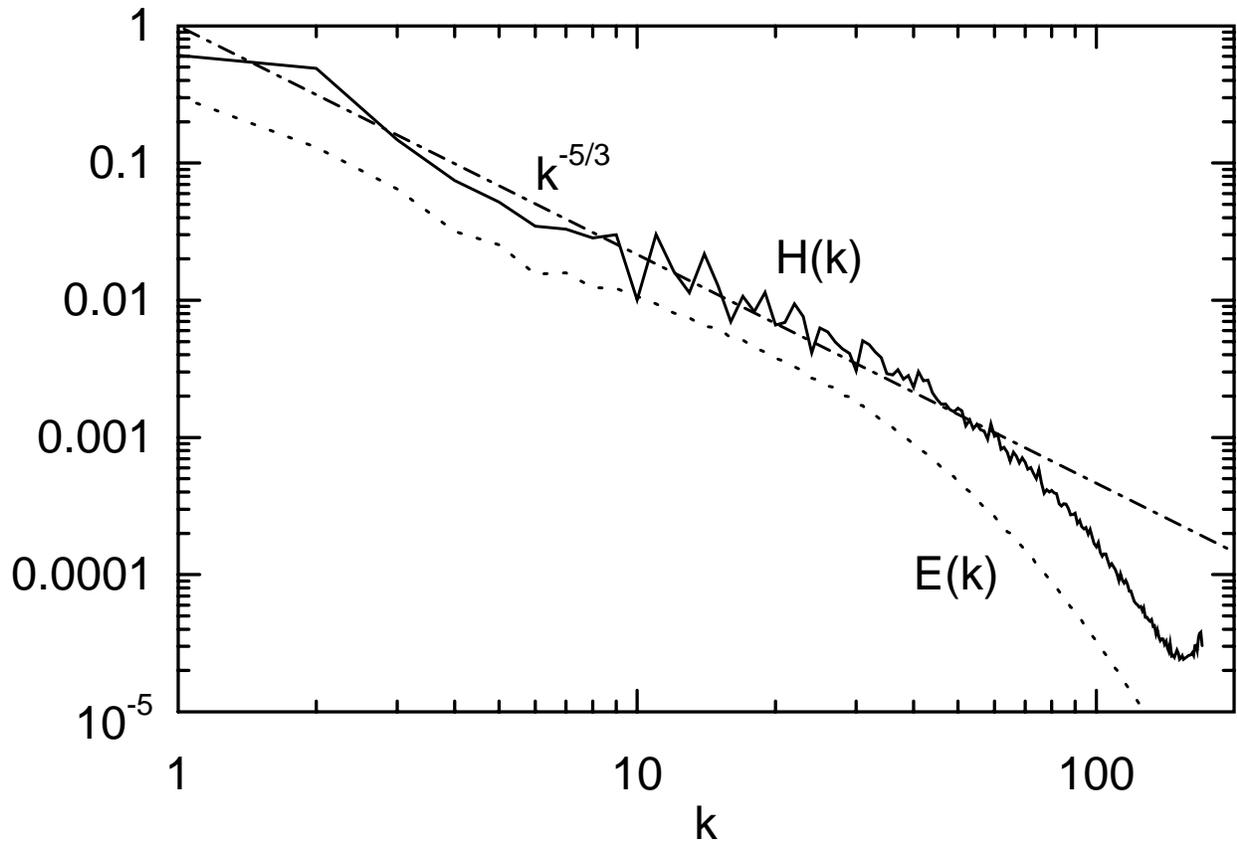
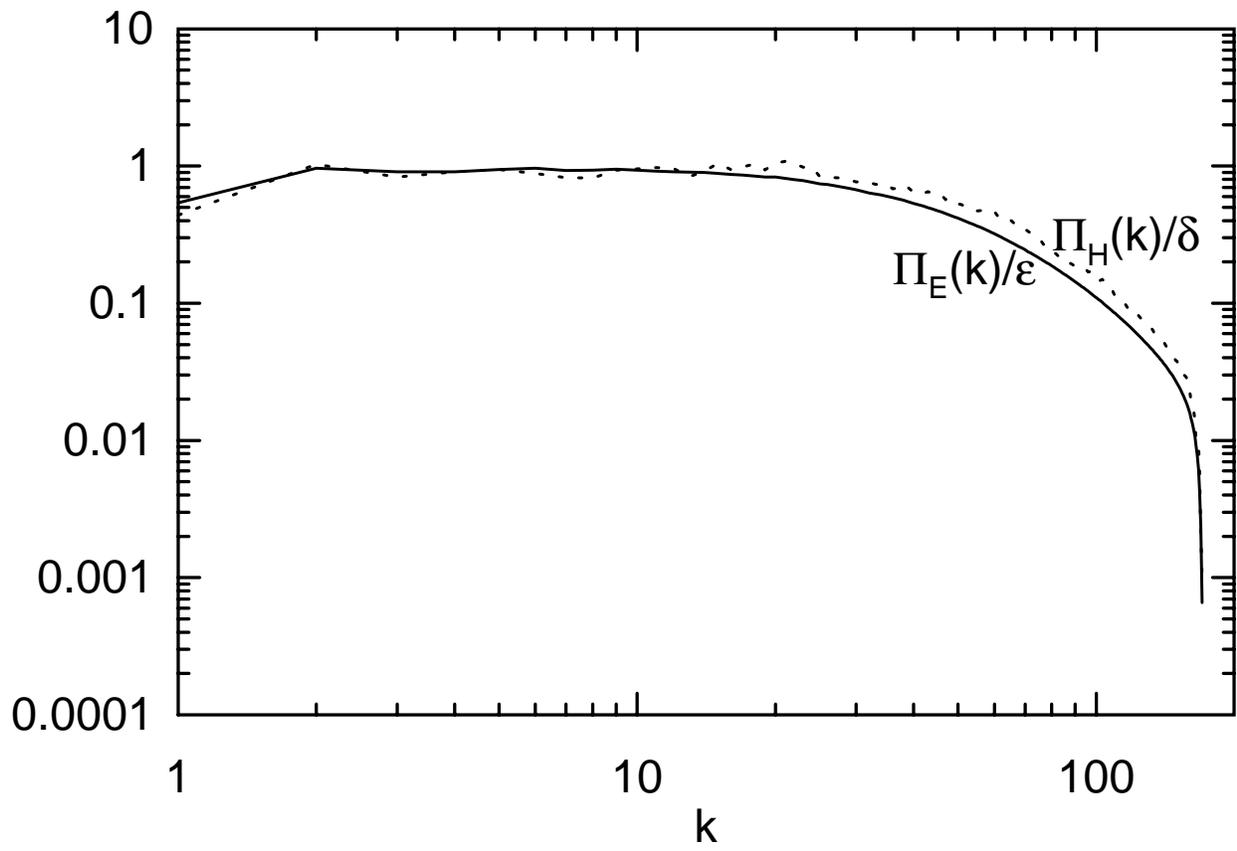
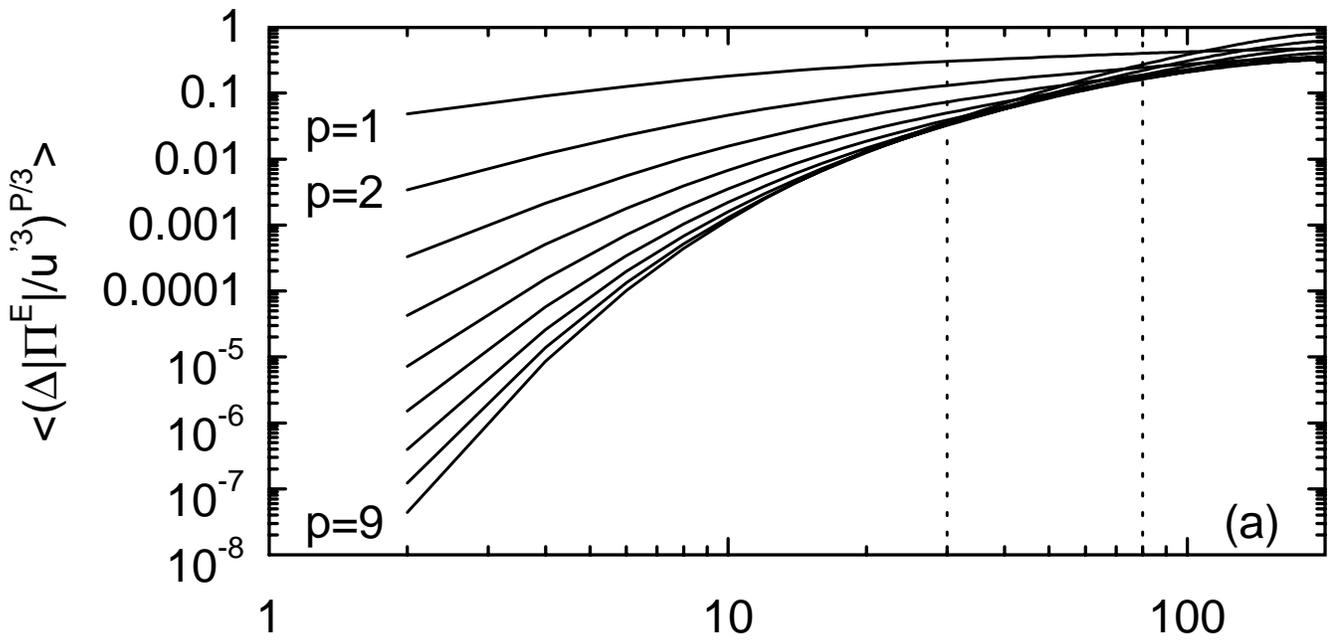
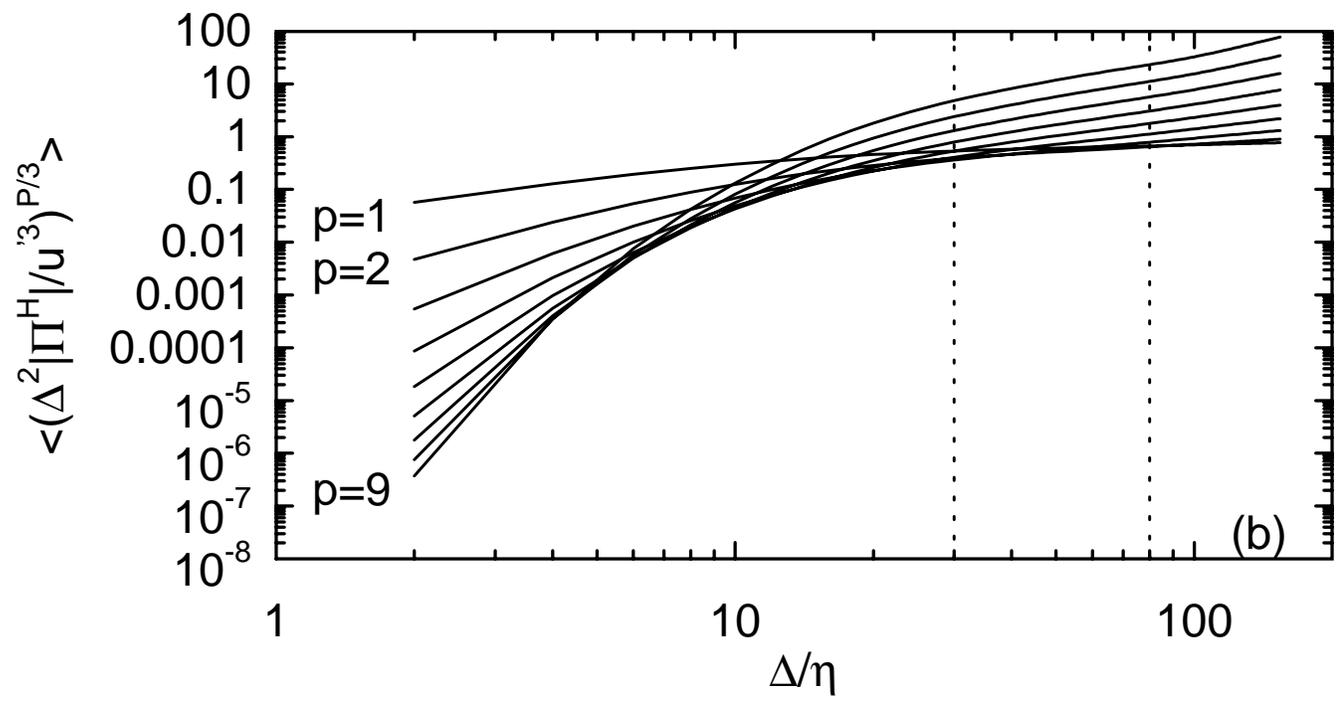
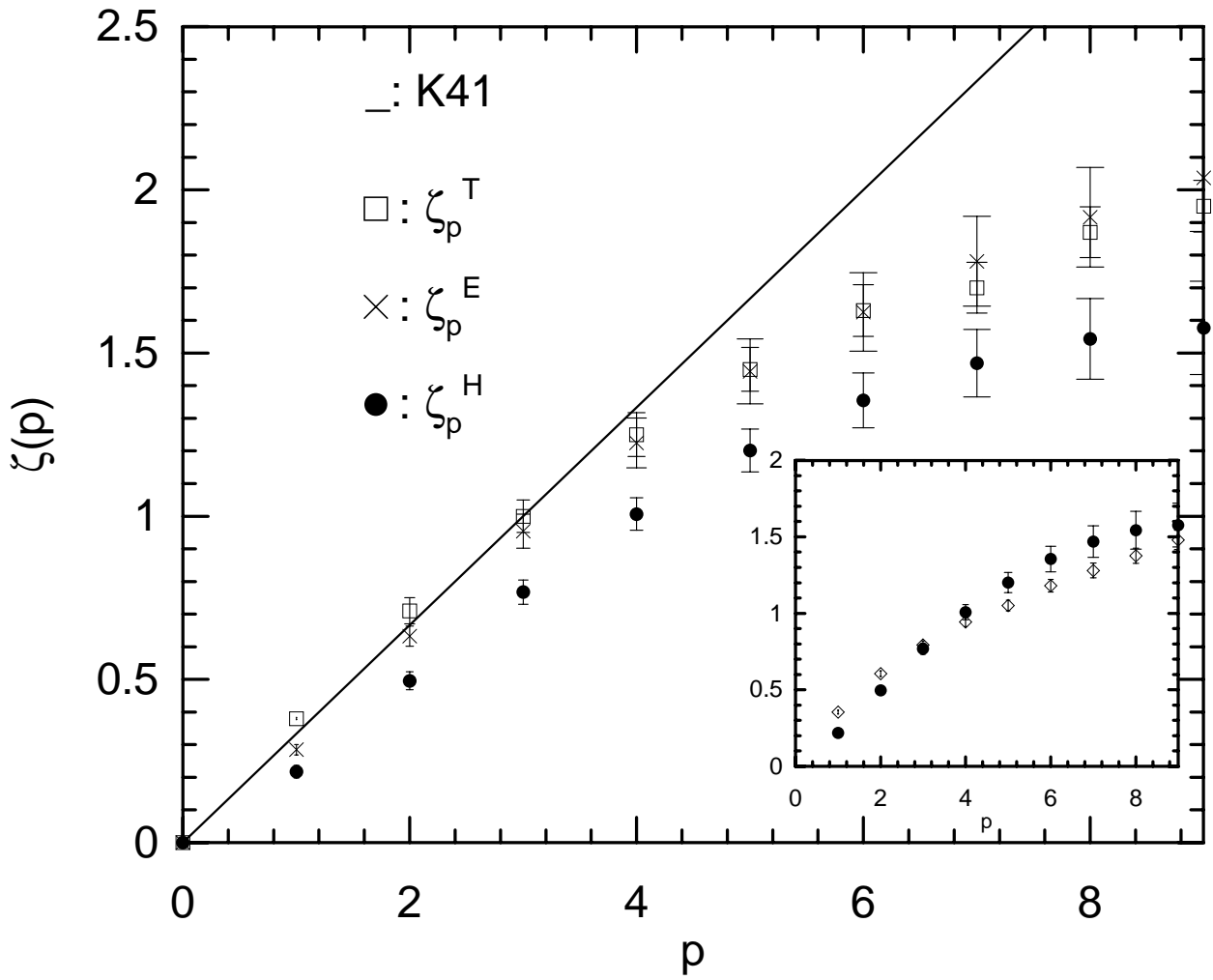
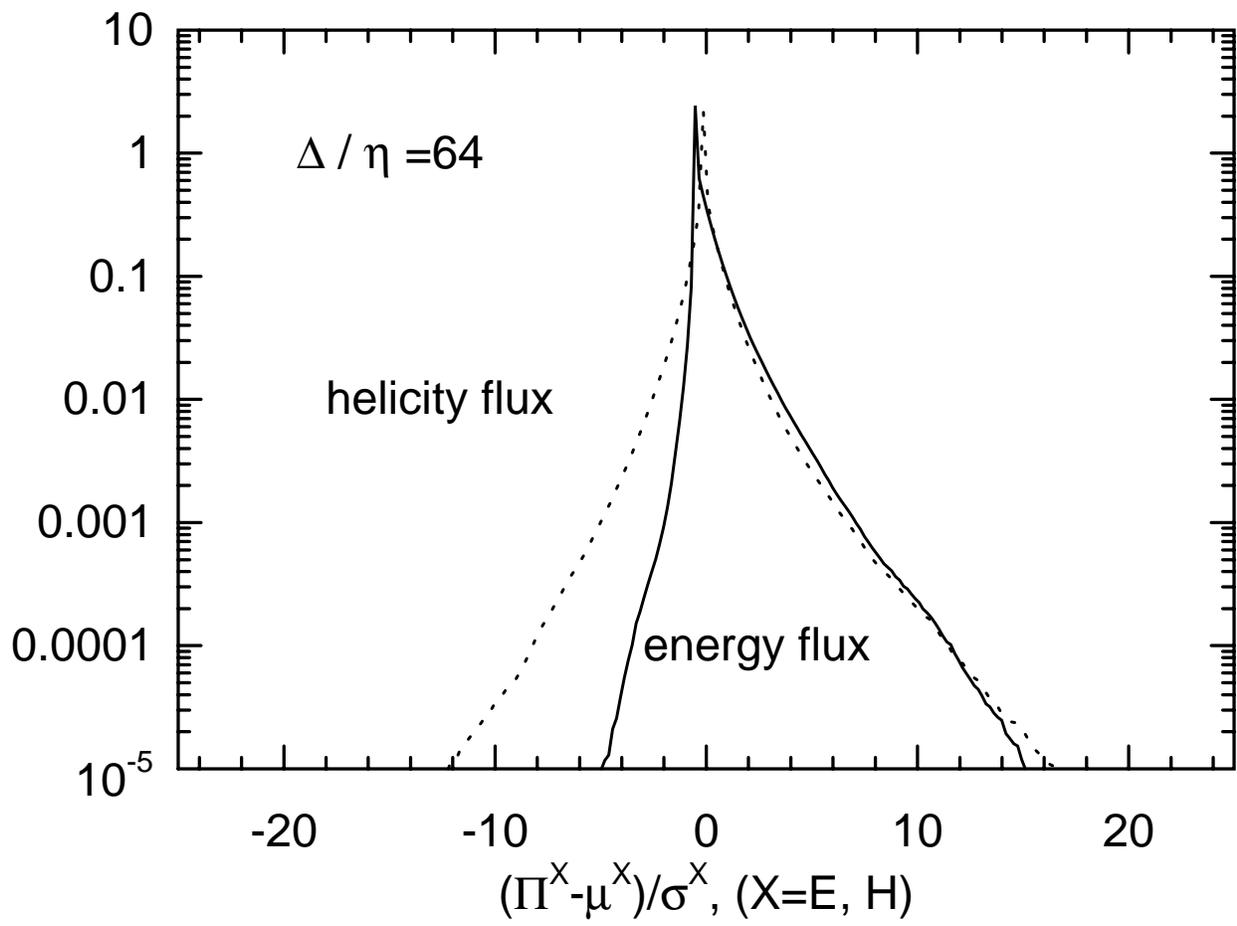
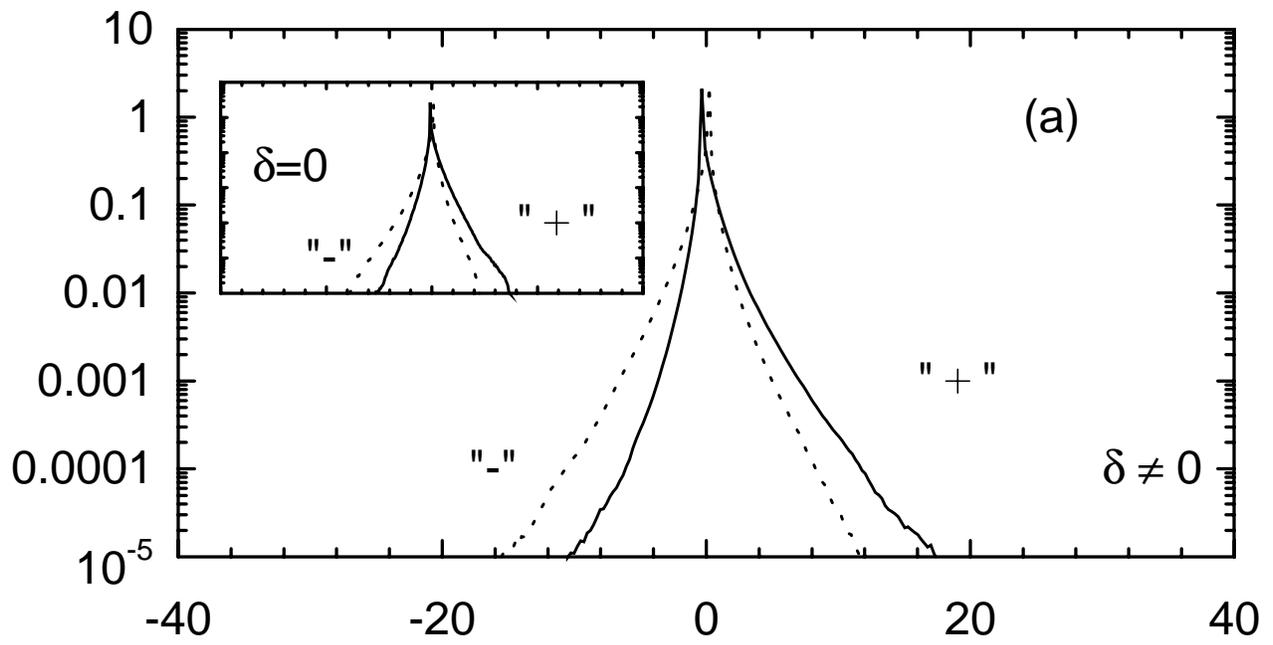
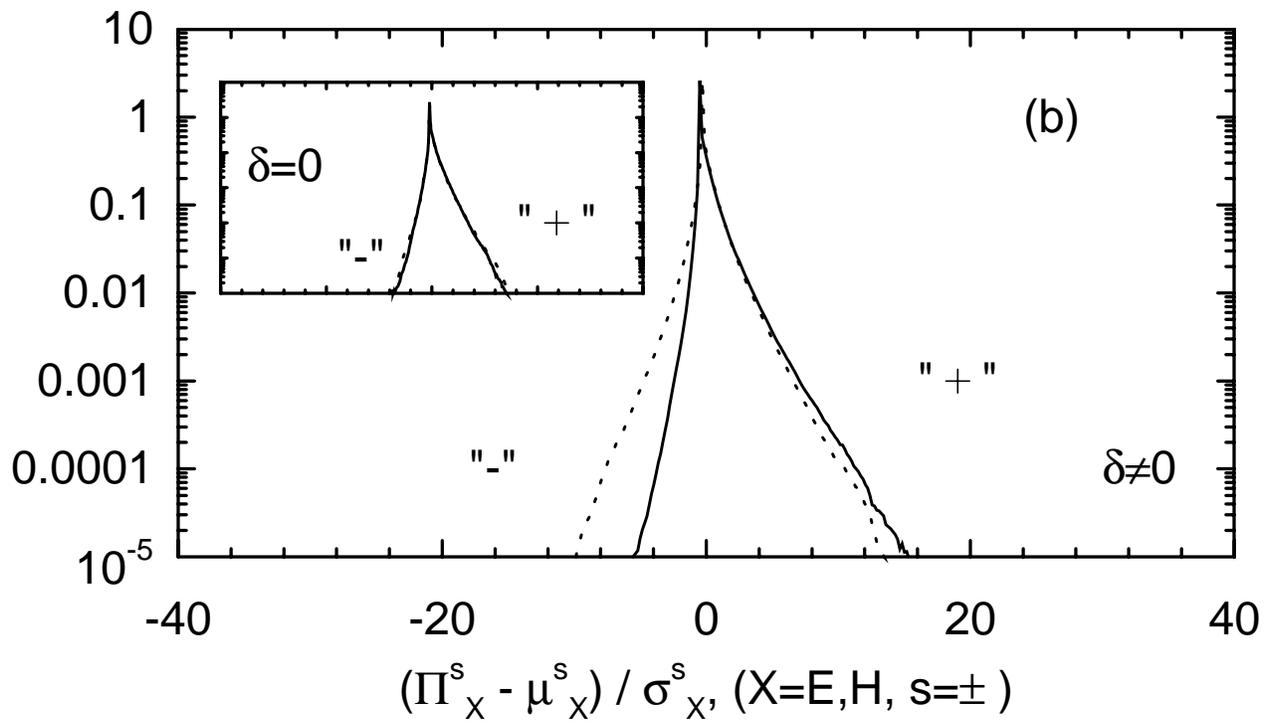
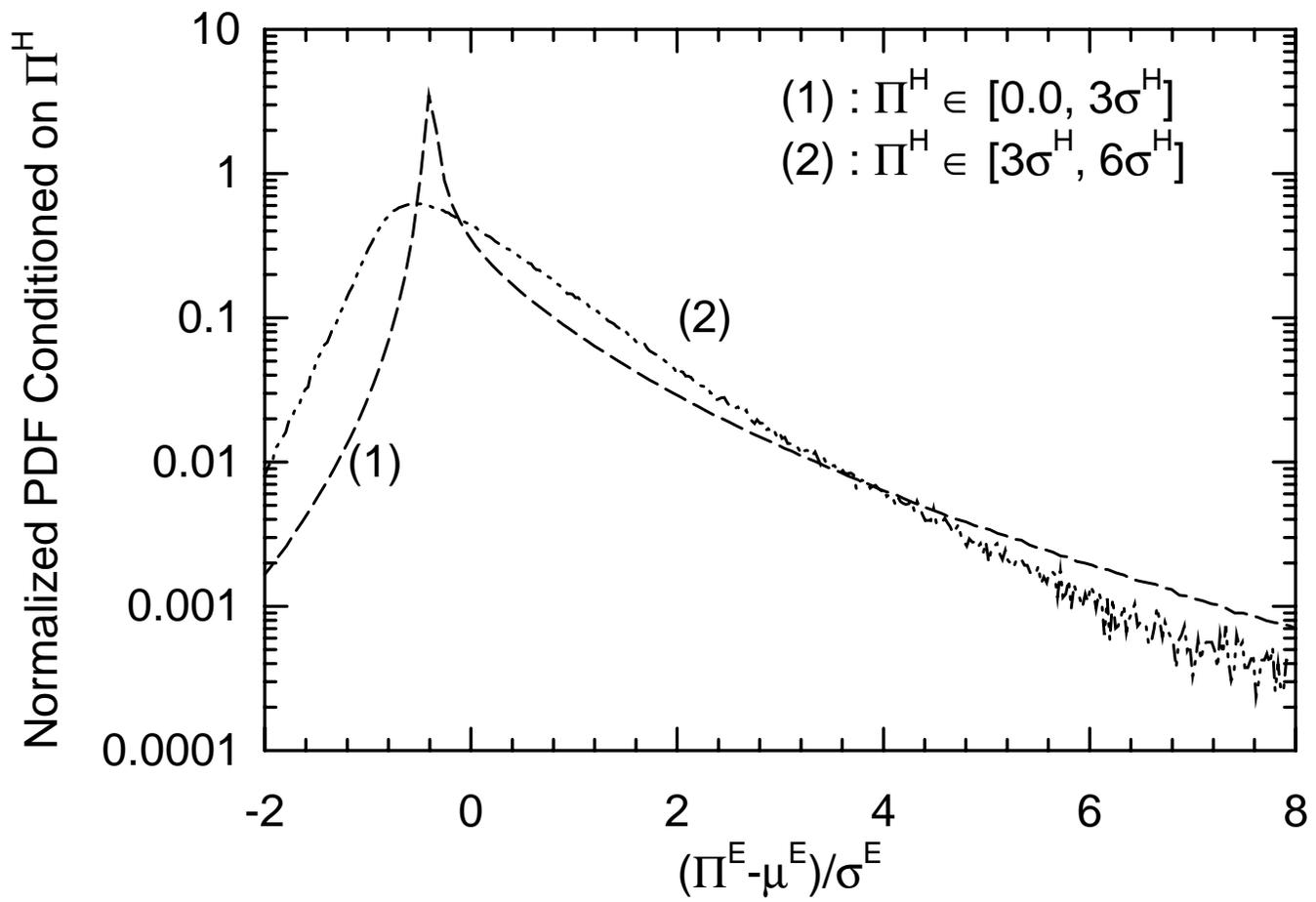
\end{document}